\documentclass[aps,prl,twocolumn,floatfix,superscriptaddress,10pt]{revtex4-2}
\usepackage{amsfonts}
\usepackage{amssymb}
\usepackage{amsmath}
\usepackage{graphicx}
\usepackage{bm}
\usepackage{color}
\usepackage{xcolor}
\usepackage{hyperref}
\hypersetup{
    colorlinks=true,
    linkcolor=blue,
    citecolor=blue,
    urlcolor=blue
}

\begin{document}

\title{Asymmetric quantum Hall effect and diminished $\nu=0$ longitudinal resistance in graphene/InSe heterostructures}

\author{Wenxue He}
\affiliation{Center for Joint Quantum Studies, Department of Physics, School of Science, Tianjin University, Tianjin 300350, China}
\affiliation{Tianjin Key Laboratory of Low Dimensional Materials Physics and Preparing Technology, School of Science, Tianjin University, Tianjin 300072, China}
\author{Shijin Li}
\affiliation{Center for Joint Quantum Studies, Department of Physics, School of Science, Tianjin University, Tianjin 300350, China}
\author{Jinhao Cheng}
\affiliation{State Key Laboratory of Advanced Materials for Intelligent Sensing, Key Laboratory of Organic Integrated Circuit, Ministry of Education \& Tianjin Key Laboratory of Molecular Optoelectronic Sciences, Department of Chemistry, School of Science \& Institute of Molecular Aggregation Science, Tianjin University, Tianjin 300072, China}
\affiliation{Collaborative Innovation Center of Chemical Science and Engineering (Tianjin), Tianjin 300072, China}
\author{Yingpeng Zhang}
\affiliation{Center for Joint Quantum Studies, Department of Physics, School of Science, Tianjin University, Tianjin 300350, China}
\affiliation{Tianjin Key Laboratory of Low Dimensional Materials Physics and Preparing Technology, School of Science, Tianjin University, Tianjin 300072, China}
\author{Kaixuan Fan}
\affiliation{Center for Joint Quantum Studies, Department of Physics, School of Science, Tianjin University, Tianjin 300350, China}
\author{Jiabo Liu}
\affiliation{Center for Joint Quantum Studies, Department of Physics, School of Science, Tianjin University, Tianjin 300350, China}
\author{Shuaishuai Ding}
\affiliation{State Key Laboratory of Advanced Materials for Intelligent Sensing, Key Laboratory of Organic Integrated Circuit, Ministry of Education \& Tianjin Key Laboratory of Molecular Optoelectronic Sciences, Department of Chemistry, School of Science \& Institute of Molecular Aggregation Science, Tianjin University, Tianjin 300072, China}
\affiliation{Collaborative Innovation Center of Chemical Science and Engineering (Tianjin), Tianjin 300072, China}
\author{Wenping Hu}
\affiliation{State Key Laboratory of Advanced Materials for Intelligent Sensing, Key Laboratory of Organic Integrated Circuit, Ministry of Education \& Tianjin Key Laboratory of Molecular Optoelectronic Sciences, Department of Chemistry, School of Science \& Institute of Molecular Aggregation Science, Tianjin University, Tianjin 300072, China}
\affiliation{Collaborative Innovation Center of Chemical Science and Engineering (Tianjin), Tianjin 300072, China}
\affiliation{The International Joint Institute of Tianjin University, Fuzhou, Tianjin University, Tianjin 300072, China}
\author{Fan Yang}
\affiliation{Center for Joint Quantum Studies, Department of Physics, School of Science, Tianjin University, Tianjin 300350, China}
\affiliation{Tianjin Key Laboratory of Low Dimensional Materials Physics and Preparing Technology, School of Science, Tianjin University, Tianjin 300072, China}
\author{Chen Wang}
\affiliation{Center for Joint Quantum Studies, Department of Physics, School of Science, Tianjin University, Tianjin 300350, China}
\author{Qing-Feng Sun}
\affiliation{International Center for Quantum Materials, School of Physics, Peking University, Beijing, China}
\author{Hechen Ren}
\email[Corresponding author: ]{ren@tju.edu.cn}
\affiliation{Center for Joint Quantum Studies, Department of Physics, School of Science, Tianjin University, Tianjin 300350, China}
\affiliation{Tianjin Key Laboratory of Low Dimensional Materials Physics and Preparing Technology, School of Science, Tianjin University, Tianjin 300072, China}
\affiliation{The International Joint Institute of Tianjin University, Fuzhou, Tianjin University, Tianjin 300072, China}


\begin{abstract}
We investigate quantum transport in graphene/InSe heterostructures and find major asymmetries in the longitudinal resistance ($R_{xx}$) and vanishing $R_{xx}$ peaks at high magnetic fields, particularly at the charge-neutrality point. Our Landauer-Buttiker analysis and numerical simulations show that a monotonically varying density gradient combined with a full equilibration mechanism can explain these phenomena. Our results also suggest the presence of trivial long-range chiral edge current and offer a broadly applicable way to engineer transport properties in quantum Hall systems.
\end{abstract}

\maketitle
\emph{Introduction.}$-$The quantum Hall effect (QHE) is a fascinating phenomenon observed in two-dimensional electron systems subjected to strong magnetic fields. Discovered by Klaus von Klitzing in 1980 \cite{Klitzing1980}, the QHE reveals that the Hall resistance, classically proportional to the magnetic field, becomes quantized into discrete plateaux. This quantization is a macroscopic manifestation of quantum mechanics, highlighting the topological nature of electron transport in these systems. The QHE has revolutionized metrology, providing an extremely accurate resistance standard. So far, it has been observed in many materials, such as GaAs \cite{Gwinn_1989, ChengyuWang_2022}, graphene \cite{Zhang2005, Novoselov2005}, HgTe \cite{BrAne_2011}, BiSbTeSe$_2$ \cite{xu_observation_2014}, Bi$_2$Se$_3$ \cite{Koirala_2015, Koirala_2019} and Bi$_2$O$_2$Se \cite{zheliuk_quantum_2024, wang_eveninteger_2024}. Among them, the zeroth Landau level (zLL) in graphene has been a subject of intense research. Although the nature of the ground state remains a hotly debated theoretical question, experimental observations have unanimously pointed to an insulating $\nu=0$ state with a pronounced peak in the longitudinal resistance \cite{Zhang2005, Novoselov2005,Abanin2007}.
\par
Between plateaux, the Hall resistance $R_{yx}$ stops being quantized and changes from one value to another. During these transitions, bulk states participate and contribute to backscattering, resulting in $R_{xx}$ peaks symmetric upon reversing the magnetic field direction. Attention given to the height of the $R_{xx}$ peaks is sparse compared to the quantized $R_{yx}$. Yet, asymmetry in longitudinal resistance is ubiquitous across material platforms. Some are by design, such as in a curved two-dimensional electron system \cite{Vorobev2007} or one with a density gradient from asymmetric positioning of evaporation sources \cite{Zhou2015}. Some are unintentional and not well understood \cite{CaddenZimansky2018, Song2019}. To our knowledge, no explanation has captured asymmetric $R_{xx}$ behaviors both in magnetic field direction and on opposite edges of the sample. 

\par
In this paper, we conduct transport experiments in graphene/InSe heterostructures, a platform reported to host a giant $\nu=2$ quantum-Hall plateau \cite{Kudrynskyi_2017}. In the graphene region uniformly covered by InSe, we observe two astonishing phenomena in the longitudinal resistance. First, it develops dramatic asymmetries in reversed magnetic fields as well as between opposite sample edges. Second, $R_{xx}$ peaks disappear at high fields throughout the transitions between $\nu=2$ and $\nu=-2$ or between $\nu=\pm2$ and $\nu=\pm6$. We interpret our experimental results both using Landauer-Buttiker analysis and through tight-binding numerical simulations while including inhomogeneities in the graphene's carrier density. The diminished longitudinal resistance in the absence of quantized transverse conductance suggests a robust equilibration mechanism between mixed chiralities, supporting the presence of long-range chiral edge currents in charge-neutral graphene at high magnetic fields \cite{AharonSteinberg2021, Lou2024}.
\par

\begin{figure}[htbp]
\centering
\includegraphics[width=\linewidth]{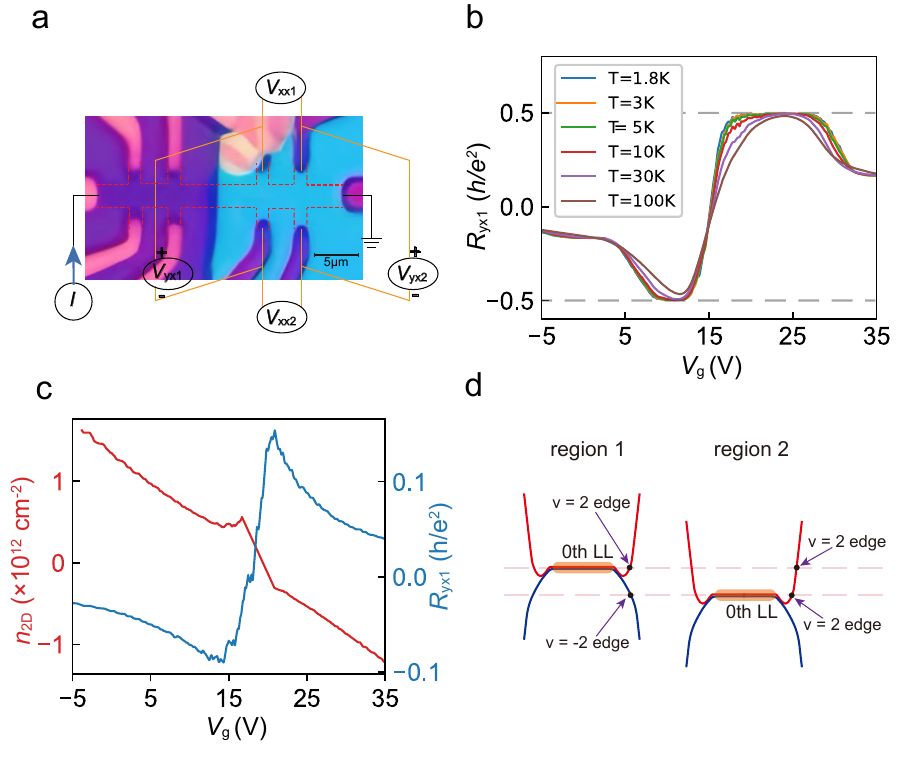}  
\caption{Graphene/InSe heterostructure Device D1. (a) Optical micrograph and measurement setup. The red dashed line outlines the graphene. (b) Hall resistance $R_{yx1}$ as a function of back-gate voltage $V_g$ measured at different temperatures. (c) Gate dependence of the calculated carrier density $n_{\mathrm{2D}}$ (red) and the corresponding $R_{yx1}$ at $B = 2\,\mathrm{T}$ (blue). (d) Trivial edge states and Landau level bending. Schematic showing the bending of the zeroth Landau level (zLL) at a trivial edge, allowing the Fermi energy ($E_F$) to create conductive edge channels while the bulk is localized.}
\label{fig_1}
\end{figure}

\emph{Charge transfer between InSe and graphene and persistent $\nu=2$ plateau.}$-$We fabricated our devices by exfoliating graphene flakes onto Si/$\mathrm{SiO_{2}}$ substrates, depositing electrodes, and etching into Hall-bar geometries (Fig.~\ref{fig_1}a). To complete the heterostructure, we covered part or all of the Hall-bar region with $\gamma$-InSe flakes ranging from 5 nm to 20 nm in thickness using a standard dry-transfer technique in a glovebox filled with argon gas. Graphene monolayers were confirmed via a combination of optical contrast, atomic force microscopy (AFM), and the sequence of quantum Hall resistance steps. Fig.~\ref{fig_1}a shows the optical image of the finished device D1, while Fig. S1 \cite{Supp} presents images of three other devices (D2, D3, and D4).
\par

Transport measurement was conducted at a base temperature of 1.8 K using circuitry shown in Fig.~\ref{fig_1}a. We used p-doped Si substrate as the backgate. As seen from the blue trace in Fig. S2b \cite{Supp}, our graphene is heavily p-doped by the $\mathrm{SiO_{2}}$ substrate, consistent with previous observations. Thanks to the charge transfer process between $\gamma$-InSe and graphene, the graphene/InSe heterostructure exhibits ambipolar conductance with charge neutrality accessible well under 100 V of backgate voltage (orange trace in Fig. S2b \cite{Supp}), lower than the previous experimental result on chemical-vapor-deposition (CVD) graphene \cite{Kudrynskyi_2017}. Most of our measurements were performed on the part of the devices covered by InSe, and the uncovered section of p-doped graphene serves as part of the source electrode.
\par

\begin{figure}[htbp]
\centering
\includegraphics[width=\linewidth]{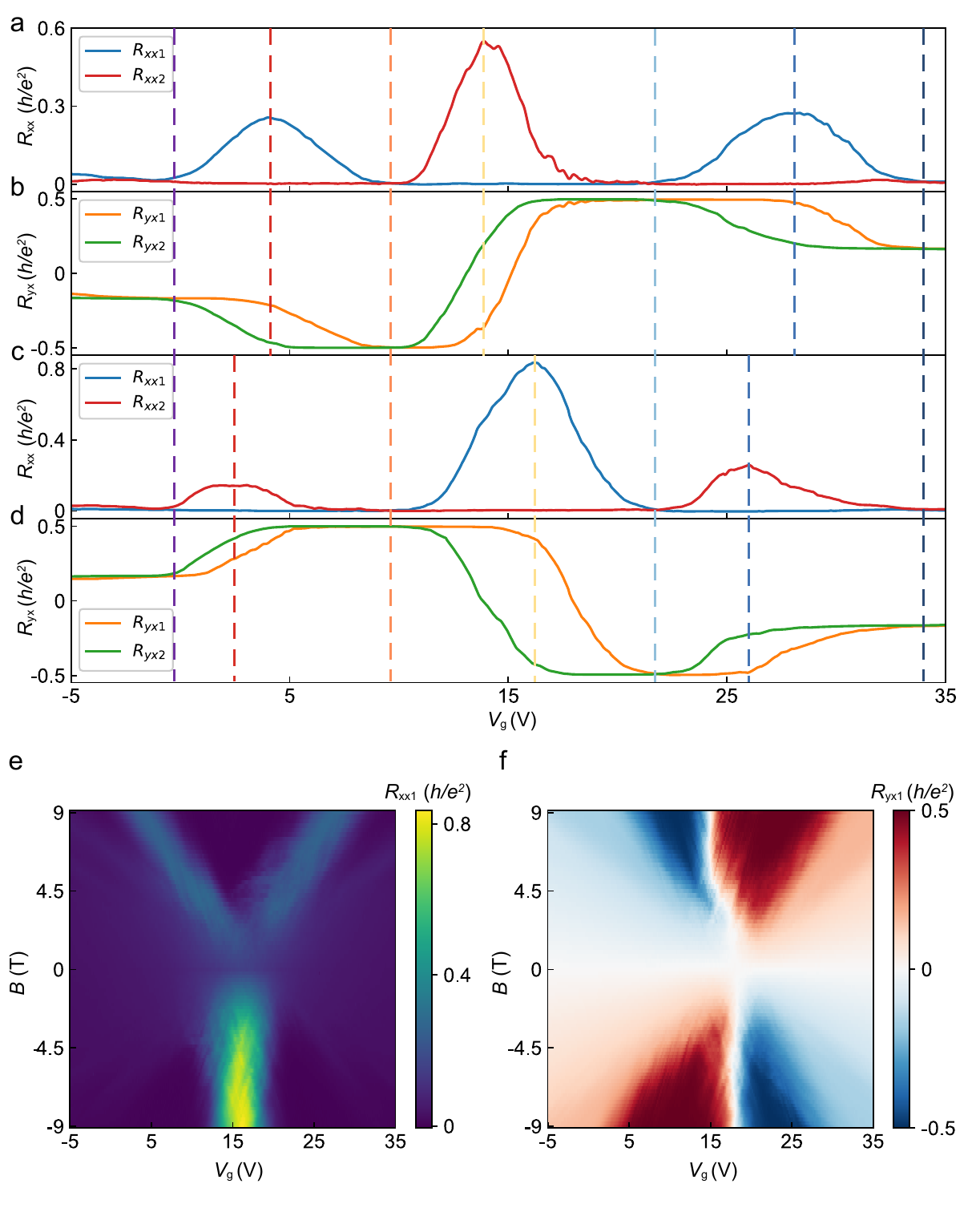}  
\caption{Magnetotransport data from device D1. (a) Longitudinal resistances ($R_{xx1}$, $R_{xx2}$) at $B = +9\,\mathrm{T}$. (b) Corresponding Hall resistances ($R_{yx1}$, $R_{yx2}$) at $B = +9\,\mathrm{T}$. (c) Longitudinal resistances ($R_{xx1}$, $R_{xx2}$) at $B = -9\,\mathrm{T}$. (d) Corresponding Hall resistances ($R_{yx1}$, $R_{yx2}$) at $B = -9\,\mathrm{T}$. (e, f) Landau fan diagrams mapping (e) $R_{xx1}$ and (f) $R_{yx1}$ as functions of back-gate voltage $V_g$ and magnetic field $B$.}
\label{fig_2}
\end{figure}

Like in CVD graphene, measurements on device D1 revealed robust $\nu=\pm2$ plateaux in the transverse resistance, observed both in magnetic field (Fig. S5c \cite{Supp}) and in gate voltage (Fig.~\ref{fig_1}b). As the temperature rises, the quantized resistance on both sides persist up to 100 K with only minor resistance deviation (Fig.~\ref{fig_1}b). Notice the $\nu=2$ plateau is wider than the $\nu=-2$ plateau, with the former spanning around 10 V in gate voltage and the latter only 5 V. At first glance, this asymmetry may seem consistent with the charge-transfer process ``pinning'' the Fermi level in the graphene at the electron side. However, we conclude otherwise based on two pieces of evidence from this device. First, the extracted carrier density as a function of gate voltage displays decent linearity with the same slope on both the electron and the hole side; it only flattens near charge neutrality where quantum capacitance dominates over geometric capacitance (Fig.~\ref{fig_1}c). This suggests in our graphene/InSe heterostructure, the graphene starts with a lower work function so the charge transfer stays roughly constant throughout our gating range, never reaching inside InSe's conduction band. A minor difference between naturally exfoliated and CVD graphene, this results in a significant consequence in the doping behavior, enabling us to access the full range of filling factors from $\nu=-6$ to $\nu=6$. Besides, the Fermi level remaining inside InSe's bandgap also explains why we measure negligible parallel conductance (evident from the quantized $R_{yx}$ values and the corresponding zero $R_{xx}$). The second piece of evidence is that when we reversed the magnetic field direction, we observed the wider plateau appearing on the hole side (Fig.~\ref{fig_2}b, d, f, S3b, and S4b \cite{Supp}). This restored symmetry in both reversing magnetic field direction and switching carrier type is prevalent throughout our study and hints at the dependence on chiral variations rather than global tunability of the Fermi level.
\par

\emph{Asymmetric quantum Hall effect and missing longitudinal resistance peaks.}$-$When the Fermi level lies between Landau levels of a quantum Hall insulator, its mesoscopic conductance follows the Landauer-Buttiker formula given by chiral edge modes \cite{Buttiker1988}. In the standard six-probe Hall-bar geometry, this means the resistance between adjacent longitudinal voltage probes drops to zero while the resistance between transverse probes yields quantized unit fractions of the von Klitzing constant. In the case of monolayer graphene, the sequence of such unit fractions reads $\pm1/2$, $\pm1/6$, etc., per the four-fold spin-valley degeneracy. When the density is tuned into the delocalized tail of a broadened Landau level, interior orbits participate in electronic transport, allowing backscattering and driving the longitudinal resistance away from zero and the transverse resistance from the quantized values. Due to this shared underlying mechanism, these two deviations normally coincide, resulting in an $R_{xx}$ peak bridging two neighboring $R_{yx}$ plateaux. In particular, the $R_{xx}$ peak in the zLL of graphene usually appears the highest of all, as shown in all previous works\cite{Zhang2005, Novoselov2005}.
\par

In our graphene/InSe heterostructures, we observe strong violations of this common behavior. When focusing on one pair of longitudinal voltage probes, we find that the longitudinal resistance becomes highly asymmetric under opposite magnetic field polarities. As shown in Fig.~\ref{fig_2}a, at $ B = +9$ T, the central peak at charge neutrality is strongly suppressed, whereas the side peaks between $\nu = \pm 2$ and $\nu = \pm 6$ remain visible. In contrast, at $ B = -9$ T (Fig.~\ref{fig_2}c), the central peak rises prominently to over $0.8\,h/e^{2}$, marking the transition between the $\nu = \pm 2$ plateaux, while the side peaks between $\nu = \pm 2$ and $\nu = \pm 6$ are absent. The corresponding Hall responses at $B = \pm 9$ T are shown in Fig.~\ref{fig_2}b and Fig.~\ref{fig_2}d, respectively. The full evolution of both longitudinal and Hall resistances with magnetic field is captured in the Landau fan diagrams in Fig.~\ref{fig_2}e, f.
\par

The intrigue is compounded by the fact that when we study the pair of voltage probes on the opposite edge, the longitudinal-resistance behavior is reversed. Fig.~\ref{fig_2}a and c compare the two sets of $R_{xx}$ as functions of gate voltage at $B = 9$ T and $B = -9$ T respectively. At $B = 9$ T, the central peak is missing in $R_{xx1}$ and the side peaks present, while the side peaks are suppressed and the central peak prominent in $R_{xx2}$ on the other side of the sample. At $B = -9$ T, the exact reverse is true, with the longitudinal resistance mimicking the behavior on the opposite side in reversed field.
\par

The discrepancy between the two $R_{xx}$'s demands corresponding discrepancy between the two $R_{yx}$'s due to Kirchhoff's Voltage Law in a closed circuit. 
\begin{equation}
R_{xx1} - R_{xx2} = R_{yx1} - R_{yx2}
\end{equation}
\par

Fig.~\ref{fig_2}b and Fig.~\ref{fig_2}d show the corresponding Hall resistances at $B = 9$ T and $B = -9$ T, respectively. The two pairs of Hall resistances in the same field appear qualitatively equivalent except for an offset in gate voltage. This offset is a strong indication of a spatially inhomogeneous density profile with a higher overall density on the right side of the sample.
\par

\begin{figure*}[htbp]
\centering
\includegraphics[width=0.8\textwidth]{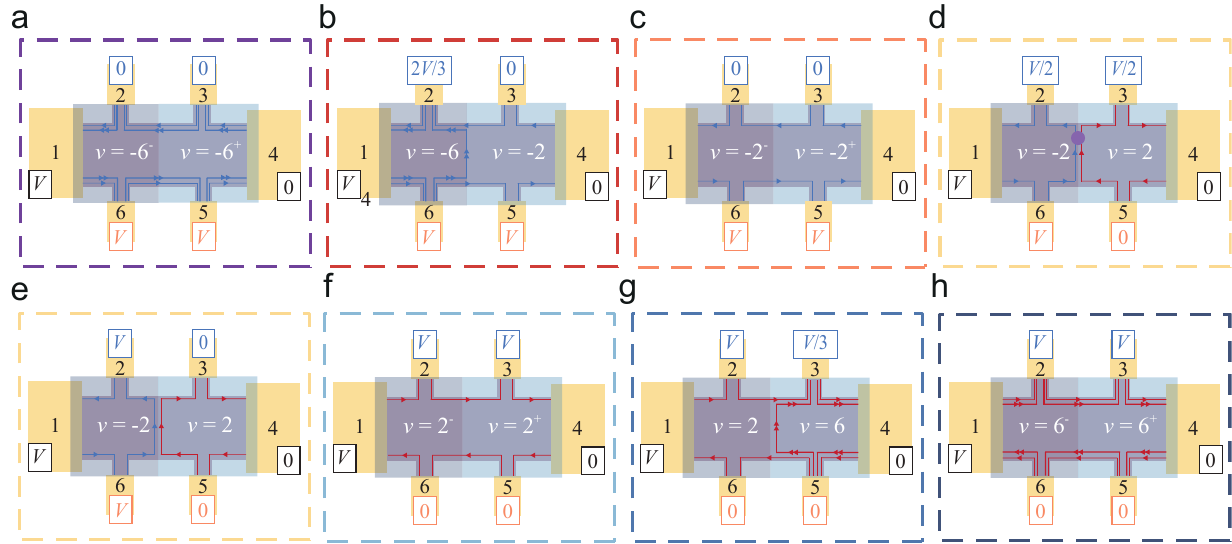}  
\caption{Landauer-Buttiker analysis of two regions with mismatched densities. (a) Global $\nu = -6$ state with insulating bulk and chiral edge states. (b) Mixed $\nu = -6$ and $\nu = -2$ phase with an additional interior edge channel. (c) Global $\nu = -2$ state. (d) Fully equilibrated counter-propagating edge channels at the interface between $\nu = -2$ and $\nu = 2$ regions. (e)Non-equilibrated channels between $\nu = -2$ and $\nu = 2$ regions preserve original potentials, yielding longitudinal voltage equal to source-drain voltage. (f) Global $\nu = 2$ state. (g) Mixed $\nu = 2$ and $\nu = 6$ phase. (h) Global $\nu = 6$ state.}
\label{fig_3}
\end{figure*}

\emph{Interpretation combining spatial inhomogeneity and non-topological edge state.}$-$We first build a toy model for Landauer-Buttiker analysis by dividing the InSe-covered graphene into two rectangular regions with slightly mismatched densities. When both regions are in the $\nu = -6$ filling (Fig.~\ref{fig_3}a), the Fermi level lies in the gap between the $N=-2$ and $N=-1$ Landau levels, the bulk is insulating, while four $N=-1$ and two $N=0$ edge states circulate the shared outer edge of the regions.
\par

As the Fermi level rises, the right region first enters the $\nu = -2$ filling (Fig.~\ref{fig_3}b), creating an interior $N=-1$ channel. A finite voltage develops between Leads 2 and 3 while Leads 5 and 6 remain equipotential, rendering a finite $R_{xx}$ on one side while keeping the other side's $R_{xx}$ zero. When we reverse the direction of the magnetic field, all the edge channels reverse their directions, swapping the longitudinal voltages on the two sides. Continue to raise the Fermi level, and we enter global $\nu = -2$ (Fig.~\ref{fig_3}c), $\nu = 2$ (Fig.~\ref{fig_3}f), and $\nu = 6$ (Fig.~\ref{fig_3}h) states, which are single-filling scenarios similar to the global $\nu = -6$ case. The mixed $\nu = 2$ and $\nu = 6$ scenario is mirror-symmetric to the $\nu = -6$ and $\nu = -2$ case.
\par

Near graphene's charge neutrality, the sample hosts two regions of $\nu = -2$ and $\nu = 2$ quantum Hall insulators, like in a graphene p-n junction \cite{Williams2007, Qingfeng_2008, Qingfeng_2010, Klimov2015, Matsuo2015}. At the interface, electron-like and hole-like edge channels co-propagate in the same direction. If they are close enough spatially, they can equilibrate fully and exit both with the averaged potential \cite{Williams2007}. This results in a zero upper $R_{xx}$ only (Fig.~\ref{fig_3}d). On the contrary, if the co-propagating channels are sufficiently separated in space, they each maintain their original potential \cite{Klimov2015}. Then, the longitudinal voltage equals the source-drain voltage on both sides (Fig.~\ref{fig_3}e). Therefore, the asymmetry bewteen $R_{xx1}$ and $R_{xx2}$ provides a gauge on the extent of the equilibration between interior edge channels of opposite chiralities. 
\par

We can also see field dependence of the charge neutrality point (CNP) in our device.  In Fig.~\ref{fig_2}f and S4 \cite{Supp}, the CNP (marked by white color) moves in gate voltage as the magnetic field increases, especially in the positive range. The equilibration picture provides a valuable perspective for understanding the offset of the CNP. Assume zero average carrier density in the whole heterostructure region corresponds to the CNP at low fields (about 17 V in gate voltage). At high fields where quantum Hall states are well developed (Fig.~\ref{fig_3}d), the right set of probes $R_{xx1}$ still exhibit half of the source-drain voltage instead of 0 V, thus shifting the CNP to the left (about 15 V).  This combination of inhomogeneity and equilibration of edge channels with opposite chiralities produce a deflected curve in the measured CNP as a function of magentic field instead of a straight line. If the edge channels between the n-doped and p-doped regions were unequilibrated (Fig.~\ref{fig_3}e), the measured Hall resistance would become undefined. The offset of apparent CNP in our experiments therefore indicates that channel equilibration has occurred.
\par

Notice in our $R_{yx}$ data, however, quantized $\nu = -2$ and $\nu = 2$ plateax never truly coexist as we sweep the gate voltage. Near charge neutrality, the Fermi levels on both sides reside in the zLL, which is famously an insulating state \cite{Zhang2005, Novoselov2005}. In cleaner, encapsulated graphene, electron-electron interaction can open a gap in the zLL, producing exotic ground states such as canted antiferromagnet and the Kekulé-distorted state \cite{Abanin2006, Nomura2006, Zhang2006, Kharitonov2012, Knothe2015, Kim2021}. However, in our graphene on $\mathrm{SiO_{2}}$, we likely have an unsplit zLL but still a narrow band thanks to the Atiyah–Singer index theorem \cite{Novoselov2005}. Consequently, in this regime between global $\nu = -2$ and global $\nu = 2$, one would expect a global insulating behavior. To address this issue, we turn to the recent discovery of the long-range non-topological edge states in charge-neutral graphene \cite{AharonSteinberg2021}. With such an edge conduction mechanism, presumably due to charge accumulation, it would cause some Landau levels to dip as they approaches the sample edge (Fig.~\ref{fig_1}d). Consequently, Fermi level would intercept the bent zLL on the edge. This bulk-edge decoupling mechanism enables coexisting localized bulk states and percolating edge states in the zLL, leading to the behavior depicted in Fig.~\ref{fig_3}d even without quantized Hall resistances.
\par

Curiously, our model only allows one interior channel mixing between the two sets of $R_{yx}$ probes. In a completely random disorder instance, this requisite is not guaranteed. For example, if there are more electron and hole puddles (four segments, for example, as depicted in Fig. S9 \cite{Supp}), a finite voltage drop will develop between Lead 2 and 3. The fact that we repeatedly observe zero longitudinal resistance on one side of the sample means our graphene/InSe heterostructures systematically produce monotonically varying density profiles. We suspect it is rooted in the heavily p-doped graphene region attached to our source lead, as it is the only element that consistently breaks the left-right mirror symmetry of the system. That said, it is remarkable that doping one of the contacts can yield a density gradient micrometers away from it (between the two sets of leads under the InSe) and a steep one at that (the interior edge channels must be close together to fully equilibrate). This gives the community studying van der Waals materials a powerful tool to generate mesoscopic density gradients, a goal previously pursued only by split gates \cite{Williams2007, Klimov2015, Matsuo2015} or mimicked by a curved sample with a field gradient \cite{Vorobev2007}. 
\par

\emph{Numerical tight-binding simulations.}$-$We extend our theoretical framework using numerical simulations based on the tight-binding model. We apply the non-equilibrium Green's function approach to compute the transmission coefficients. Based on the Landauer-Büttiker formalism, we can directly calculate the longitudinal and Hall resistances from the transmission coefficients. Given our experimental results indicate no spin polarization, we simplify our calculations using a spinless graphene model subjected to a uniform magnetic field. This magnetic field is introduced via a phase in the hopping terms, a technique known as Peierls' substitution~\cite{Peierls_1933, Supp}. To model the effect of InSe, we utilize an on-site term \( V_{\bm{i}} \). We consider two types of potential profiles. The first is a spatially correlated disorder potential: 

\begin{equation} 
\begin{gathered} 
V_{\bm{i}}=\sum_{\bm{j}}u_{\bm{j}}\exp(-|\bm{r}_{\bm{i}}-\bm{r}_{\bm{j}}|/\xi).
\end{gathered}\label{eq_tb_1} 
\end{equation} 
Here, \( \bm{r}_{\bm{i}}=(m_i a,n_i a) \) represents a lattice site, where \( m_i \) and \( n_i \) are integers and \( a \) is the lattice constant. The term \( u_{\bm{i}} \) is randomly and uniformly distributed within the range of \( [-W/2,W/2] \). The positive parameter \( \xi \) characterizes the correlation length of the impurities. To account for the asymmetry observed in the quantum Hall effect, we also introduce an anisotropic energy component along the \( x \)-direction, \( V_{\bm{i}}=\epsilon_0 +bm_{\bm{i}}a\), with \( \epsilon_0 ,  b  > 0\) .

\begin{figure}[htbp]
\centering
\includegraphics[width=\linewidth]{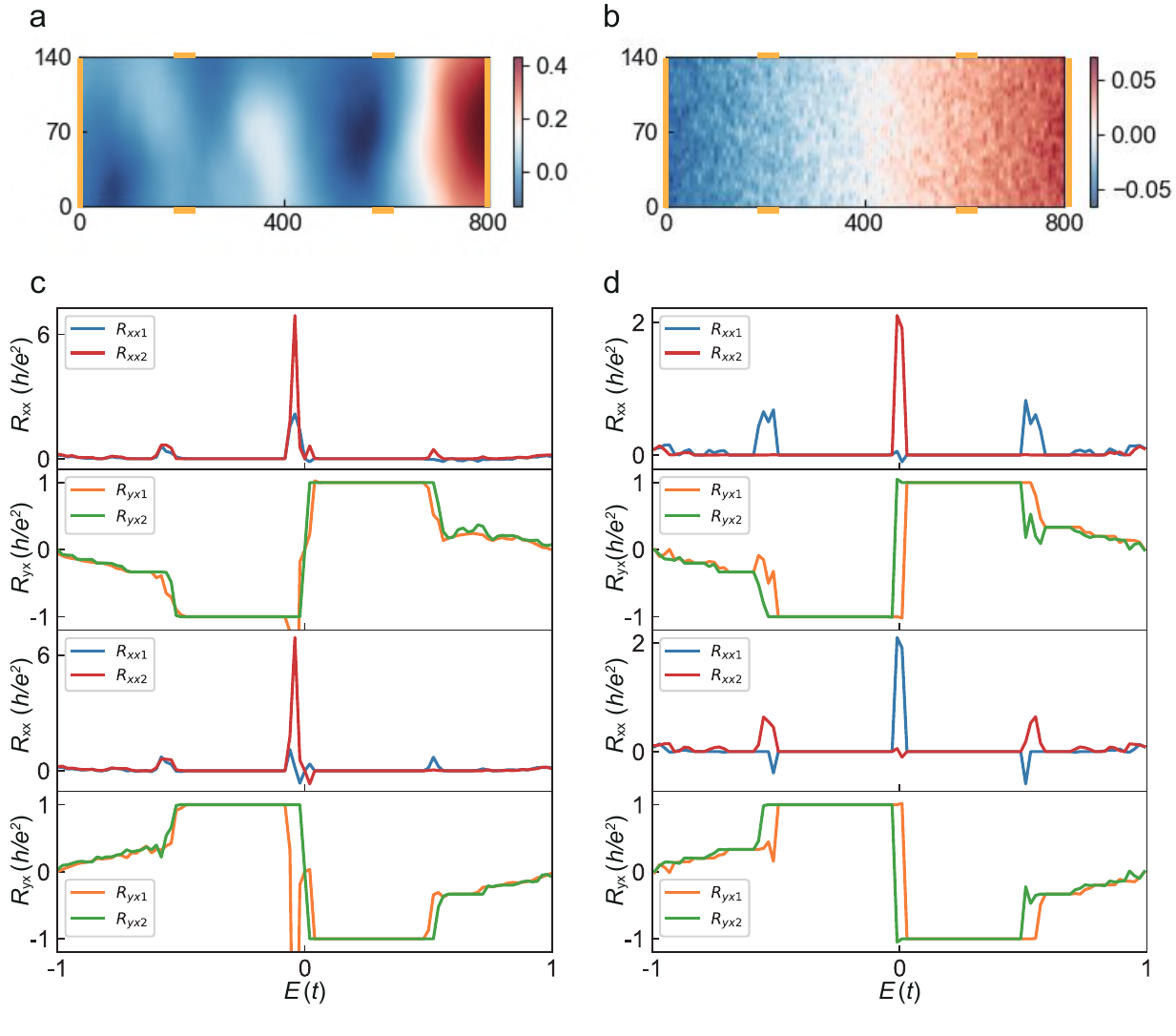}  
\caption{Tight-binding simulation of quantum Hall transport.  
(a) Chemical potential profile with Gaussian random disorder ($W=0.03$, $\xi=150a$).  
(b) Chemical potential under linear gradient (left potential $V_\text{left}=-0.05$, right potential $V_\text{right}=0.05$) with added disorder ($W=0.02$, $\xi=2a$).  
(c) Resistances for random disorder at positive (top panels) and negative (bottom panels) fields.  
(d) Resistances for gradient with disorder at positive (top panels) and negative (bottom panels) fields.  
Gold segments indicate electrodes in (a,b).}
\label{fig_4}
\end{figure} 

In both types of potential profiles, we observe significant discrepancy between $R_{xx1}$ and $R_{xx2}$, reproducing the pronounced spatial and magnetic-field asymmetry. The Gaussian-correlated disorder profile (Fig.~\ref{fig_4}a) reproduces the asymmetry but not the diminishing $R_{xx}$ peaks (Fig.~\ref{fig_4}c). In the case of a linear gradient with added disorder (Fig.~\ref{fig_4}b), $R_{xx1}$ shows no peak at charge neutrality but exhibits side transition peaks, while $R_{xx2}$ mirrors this behavior upon field reversal. This confirms that the observed field and spatial asymmetries,especially the diminished $R_{xx}$ peaks, primarily stem from the presence of a strong monotonic density gradient.

Our numerical calculations also capture the offset of the CNP. Using the chemical potential profile in Fig.~\ref{fig_4}b, we computed the evolution of $R_{yx1}$ and $R_{yx2}$ with energy $E$ at a magnetic field $B \sim 1/N$ (where $N$ is the Landau level index), as shown in Fig. S11 \cite{Supp}. We observe that under finite magnetic fields, the charge neutrality point (CNP) of $R_{yx1}$ shifts to lower energies, while that of $R_{yx2}$ shifts to higher energies. This opposing CNP offset remains robust over a wide range of magnetic fields, consistent with our experimental observations in Fig.~\ref{fig_2}f.

\emph{Conclusion.~}In conclusion, our experimental results on graphene/InSe heterostructures revealed two striking transport phenomena. We observed significant asymmetries in longitudinal resistance under reversed magnetic fields and on opposite edges of the same sample. In several devices, the longitudinal resistance peak vanished at high fields during transitions between filling factors, even at the charge-neutrality point ($\nu = 0$). Our analysis, incorporating Landauer-Buttiker theory and tight-binding simulations with carrier density inhomogeneities, suggests that this asymmetry and reduced longitudinal resistance can originate from a monotonically varying density gradient. A random distribution of Gaussian disorder with appropriate correlation lengths reproduces the asymmetry but not a zero $R_{xx}$. Without quantized $R_{yx}$ at $\nu = 0$, the diminished longitudinal resistance on one side of the sample points to a strong equilibration between well-developed chiral channels and supports the existence of long-range chiral edge currents in charge-neutral graphene under high magnetic fields. Our findings apply to all two-dimensional electron systems exhibiting quantum Hall effects and suggest a simple way to generate mesoscopic density gradient.
\par

\begin{acknowledgments}
This work is supported by the National Natural Science Foundation of China. It is also sponsored by the National Key Research and Development Program of China under Grant No. 2024YFA1209600 and the National Natural Science Foundation of China under Grant No. 52373250.
\end{acknowledgments}

\bibliography{mainrefs-V18}


\end{document}


\title{Supplemental Materials: Asymmetric quantum Hall effect and diminished $\nu=0$ longitudinal resistance in graphene/InSe heterostructures}

\author{Wenxue He}
\affiliation{Center for Joint Quantum Studies, Department of Physics, School of Science, Tianjin University, Tianjin 300350, China}
\affiliation{Tianjin Key Laboratory of Low Dimensional Materials Physics and Preparing Technology, School of Science, Tianjin University, Tianjin 300072, China}
\author{Shijin Li}
\affiliation{Center for Joint Quantum Studies, Department of Physics, School of Science, Tianjin University, Tianjin 300350, China}
\author{Jinhao Cheng}
\affiliation{State Key Laboratory of Advanced Materials for Intelligent Sensing, Key Laboratory of Organic Integrated Circuit, Ministry of Education \& Tianjin Key Laboratory of Molecular Optoelectronic Sciences, Department of Chemistry, School of Science \& Institute of Molecular Aggregation Science, Tianjin University, Tianjin 300072, China}
\affiliation{Collaborative Innovation Center of Chemical Science and Engineering (Tianjin), Tianjin 300072, China}
\author{Yingpeng Zhang}
\affiliation{Center for Joint Quantum Studies, Department of Physics, School of Science, Tianjin University, Tianjin 300350, China}
\affiliation{Tianjin Key Laboratory of Low Dimensional Materials Physics and Preparing Technology, School of Science, Tianjin University, Tianjin 300072, China}
\author{Kaixuan Fan}
\affiliation{Center for Joint Quantum Studies, Department of Physics, School of Science, Tianjin University, Tianjin 300350, China}
\author{Jiabo Liu}
\affiliation{Center for Joint Quantum Studies, Department of Physics, School of Science, Tianjin University, Tianjin 300350, China}
\author{Shuaishuai Ding}
\affiliation{State Key Laboratory of Advanced Materials for Intelligent Sensing, Key Laboratory of Organic Integrated Circuit, Ministry of Education \& Tianjin Key Laboratory of Molecular Optoelectronic Sciences, Department of Chemistry, School of Science \& Institute of Molecular Aggregation Science, Tianjin University, Tianjin 300072, China}
\affiliation{Collaborative Innovation Center of Chemical Science and Engineering (Tianjin), Tianjin 300072, China}
\author{Wenping Hu}
\affiliation{State Key Laboratory of Advanced Materials for Intelligent Sensing, Key Laboratory of Organic Integrated Circuit, Ministry of Education \& Tianjin Key Laboratory of Molecular Optoelectronic Sciences, Department of Chemistry, School of Science \& Institute of Molecular Aggregation Science, Tianjin University, Tianjin 300072, China}
\affiliation{Collaborative Innovation Center of Chemical Science and Engineering (Tianjin), Tianjin 300072, China}
\affiliation{The International Joint Institute of Tianjin University, Fuzhou, Tianjin University, Tianjin 300072, China}
\author{Fan Yang}
\affiliation{Center for Joint Quantum Studies, Department of Physics, School of Science, Tianjin University, Tianjin 300350, China}
\affiliation{Tianjin Key Laboratory of Low Dimensional Materials Physics and Preparing Technology, School of Science, Tianjin University, Tianjin 300072, China}
\author{Chen Wang}
\affiliation{Center for Joint Quantum Studies, Department of Physics, School of Science, Tianjin University, Tianjin 300350, China}
\author{Qing-Feng Sun}
\affiliation{International Center for Quantum Materials, School of Physics, Peking University, Beijing, China}
\author{Hechen Ren}
\email[Corresponding author: ]{ren@tju.edu.cn}
\affiliation{Center for Joint Quantum Studies, Department of Physics, School of Science, Tianjin University, Tianjin 300350, China}
\affiliation{Tianjin Key Laboratory of Low Dimensional Materials Physics and Preparing Technology, School of Science, Tianjin University, Tianjin 300072, China}
\affiliation{The International Joint Institute of Tianjin University, Fuzhou, Tianjin University, Tianjin 300072, China}


\begin{abstract}
These Supplemental Materials provide comprehensive experimental and theoretical support for the investigation of the asymmetric quantum Hall effect in graphene/InSe heterostructures. We present structural characterization of multiple devices (D1--D5) using atomic force microscopy and optical microscopy, revealing bubbles and wrinkles that contribute to morphological disorder. Detailed magnetotransport measurements demonstrate the robustness of the transport asymmetry under varying magnetic fields and temperatures. Temperature-dependent studies reveal distinctive thermal activation behaviors and significant gap asymmetry between opposite magnetic field polarities. Gate hysteresis measurements establish reproducible ferroelectric switching in the $\gamma$-InSe layer. Through systematic tight-binding simulations, we explore how disorder strength $W$ and correlation length $\xi$ affect transport asymmetry, showing that while random disorder produces certain asymmetric features, a monotonic density gradient is necessary to explain the most dramatic observations. These integrated studies provide multifaceted evidence supporting the interpretation presented in the main text.
\end{abstract}

\maketitle
\tableofcontents
\newpage
\renewcommand{\thefigure}{S\arabic{figure}}
\setcounter{figure}{0}   

\section*{S1: Characterization of Graphene/InSe Heterostructures}
\label{sec1}
Monolayer graphene was mechanically exfoliated onto a silicon substrate with 285 nm of thermally grown oxide layer. The graphene was subsequently annealed at 300\,\textdegree{}C for 10 hours in a mixed Ar/H\textsubscript{2} atmosphere (9:1 ratio) to remove polymeric residues. Hall-bar electrodes were then patterned via ultraviolet (UV) photolithography. A 10 nm-thick Au electrode layer was deposited by magnetron sputtering. An additional photolithography step was performed, followed by reactive ion etching (RIE) to define the Hall-bar region. Few-layer $\gamma$-InSe flakes (HQ Graphene), ranging from 5 nm to 20 nm in thickness, were exfoliated onto a polydimethylsiloxane (PDMS) stamp using a standard dry-transfer technique within an Argon-filled glovebox. The InSe flake was precisely aligned to the predefined graphene Hall-bar using a micromanipulator-equipped transfer stage. The stamp was then released by heating at 80\,\textdegree{}C for 10 minutes, yielding the graphene/InSe van der Waals heterostructure.

\begin{figure*}[htbp]
\centering
\includegraphics[width=0.8\textwidth]{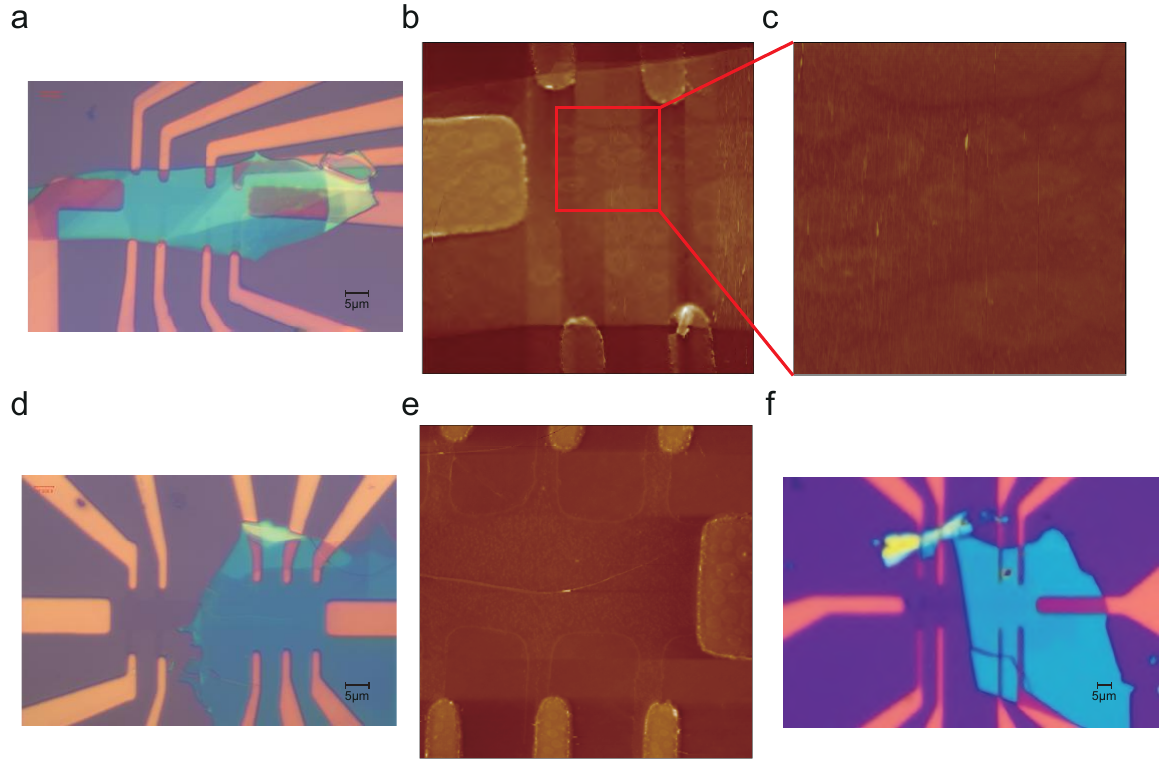}  
\caption{
\textbf{Structural characterization of graphene/InSe heterostructure devices.}
(a) Optical micrograph of device D2. (b,c) Atomic force microscopy (AFM) images of device D2 at different regions. (d) Optical micrograph and (e) corresponding AFM image of device D3. (f) Optical micrograph of device D4.}
\label{figs_1}
\end{figure*}

Structural characterization confirmed the successful fabrication of the heterostructures. It also revealed common transfer-related artifacts. Optical micrographs of devices D2, D3, and D4 are shown in Fig.~\ref{figs_1}(a, d, f). These images verify the overall device geometry and the accurate alignment of the van der Waals interfaces. We used atomic force microscopy (AFM) to further examine the surface morphology and flake thickness. The AFM images in Fig.~\ref{figs_1}(b, c, e) show localized bubbles and wrinkles. These features are not visible in optical microscopy. The bubbles have small vertical profiles but large lateral sizes, often hundreds of nanometers across. Device D3, in particular, showed prominent wrinkle networks. These structural imperfections introduce morphological disorder at the interface. This disorder may influence the local electronic environment and interfacial coupling. 
\par

\begin{figure*}[htbp]
\centering
\includegraphics[width=0.8\textwidth]{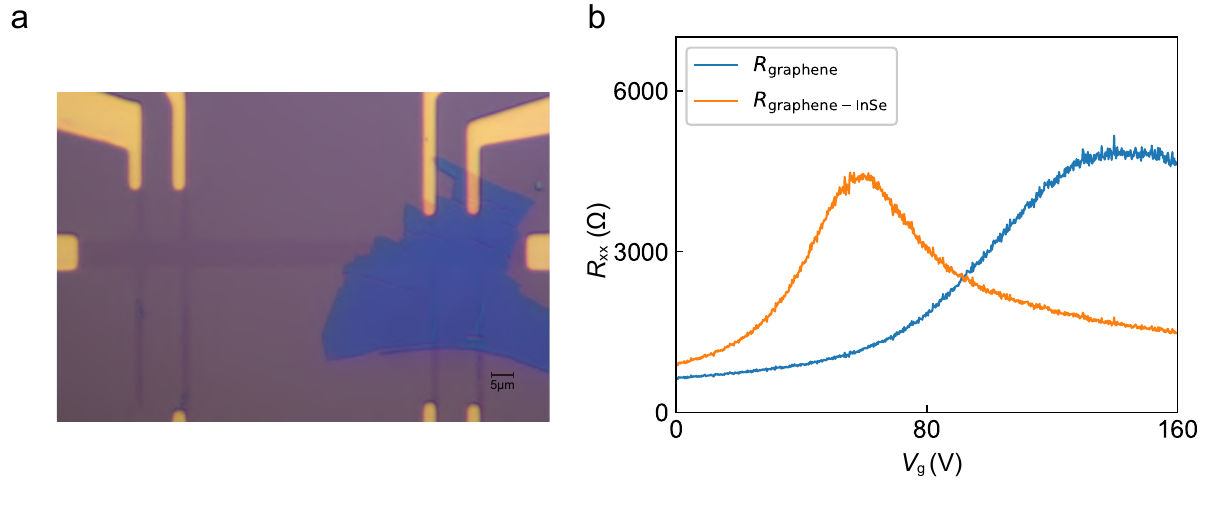}  
\caption{
\textbf{Room-temperature measurements.}
(a) Optical microscopy image of device D5. (b) Room-temperature measurements of $R_{\mathrm{graphene}}$ and $R_{\mathrm{graphene-InSe}}$ on both sides of device D5. Comparative analysis reveals charge transfer from InSe to graphene.}
\label{figs_2}
\end{figure*}

An optical micrograph of Device D5 is shown in Fig.~\ref{figs_2}a. Room-temperature measurements in Fig.~\ref{figs_2}b compare $R_{\mathrm{graphene}}$ and $R_{\mathrm{graphene-InSe}}$ across different device regions, revealing significant charge transfer from InSe to graphene.
\par

\section*{S2: Asymmetric Quantum Hall Effect in Different Devices}
\label{sec2}

Electrical transport measurements were performed using a Physical Property Measurement System (PPMS DynaCool, Quantum Design Inc.) with a magnetic field range of $\pm$9 T. The longitudinal ($R_{xx}$) and Hall ($R_{yx}$) resistances were acquired using a standard low-frequency lock-in technique. An 80 nA AC excitation was supplied by an SR830 lock-in. The resulting voltage signals were measured by a second SR830 lock-in amplifier. Backgate voltage ($V_g$) was applied to the heavily doped silicon substrate using a Keithley 2636B SourceMeter unit.

\begin{figure*}[htbp]
\centering
\includegraphics[width=0.8\textwidth]{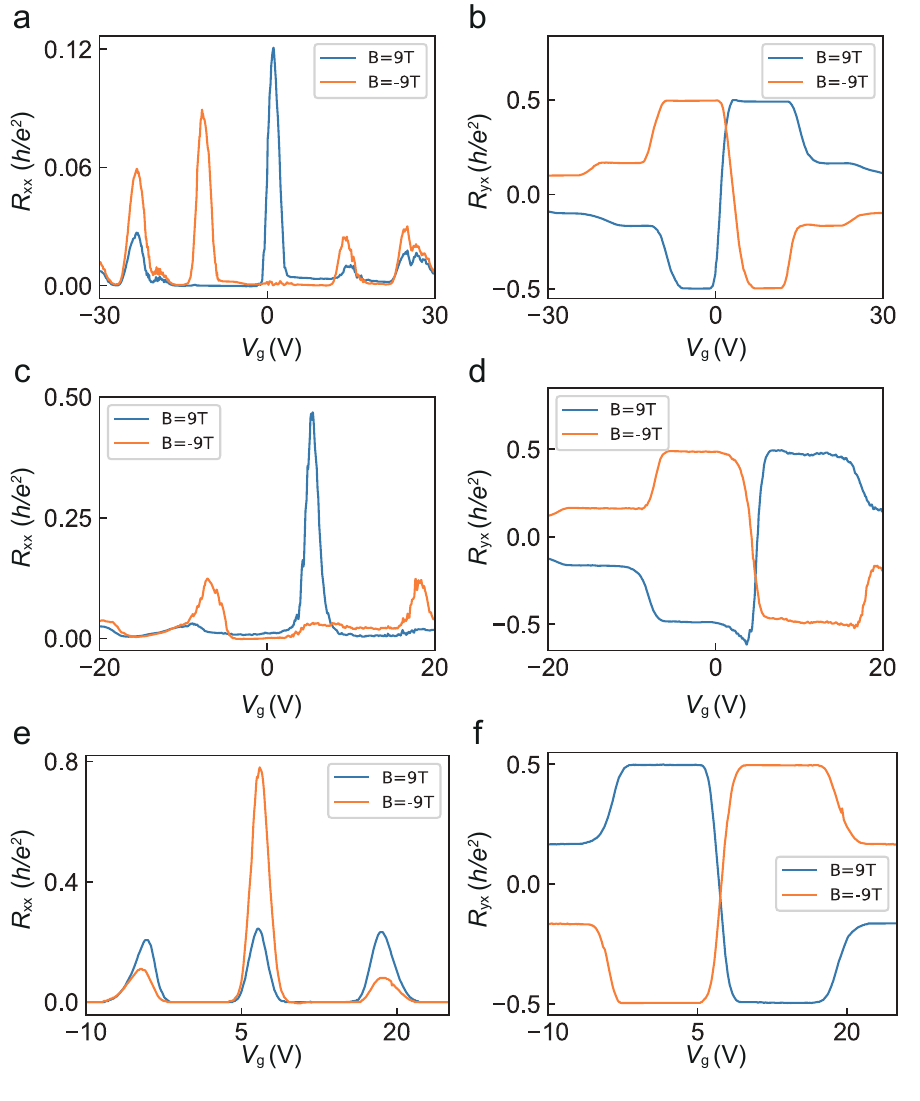}  
\caption{(a,b) Longitudinal resistance $R_{xx}$ and Hall resistance $R_{yx}$ of device D2 measured at $B = +9$ T and $B = -9$ T.  
At $B = -9$ T the $\nu=0$ longitudinal resistance is diminished and approaches zero, while at $B = +9$ T the $\nu=-2$ longitudinal resistance is likewise diminished; the plateau width of the Hall resistance differs at $\nu=+2$ and $\nu=-2$.  
(c,d) Longitudinal resistance $R_{xx}$ and Hall resistance $R_{yx}$ of device D3 at $B = +9$ T and $B = -9$ T; the $\nu=0$ longitudinal resistance is suppressed at $B = -9$ T.  
(e,f) Longitudinal resistance $R_{xx}$ and Hall resistance $R_{yx}$ of device D4 at $B = +9$ T and $B = -9$ T; the $\nu=0$ longitudinal-resistance peak at $B = +9$ T is approximately $31\%$ of that at $B = -9$ T.}
\label{figs_3}
\end{figure*}

Under these measurement conditions, we consistently observed the emergence of an asymmetric quantum Hall response across multiple devices. Fig.~\ref{figs_3} presents the magnetotransport characteristics of Devices D2, D3, and D4, measured at $B = \pm 9$ T, which collectively underscore the robustness and device-to-device variability of the effect.

In Device D2 (Fig.~\ref{figs_3} a, b), a pronounced magnetic field polarity dependence is observed. At $B = -9$ T, the longitudinal resistance $R_{xx}$ at charge neutrality ($\nu=0$) is substantially diminished and approaches zero, while at $B = +9$ T, the $\nu=-2$ longitudinal resistance is likewise diminished. The Hall resistance $R_{yx}$ exhibits a discernible asymmetry in the plateau widths between the $\nu=+2$ and $\nu=-2$ states. Device D3 (Fig.~\ref{figs_3} c, d) shows a similar trend of field-polarity-dependent suppression, with the $\nu=0$ $R_{xx}$ peak strongly suppressed at $B = -9$ T. The effect is further quantified in Device D4 (Fig.~\ref{figs_3} e, f), where the amplitude of the $\nu=0$ longitudinal resistance peak at $B = +9$ T is reduced to approximately $31\%$ of its value at the opposite magnetic field.
\par

\begin{figure*}[htbp]
\centering
\includegraphics[width=0.8\textwidth]{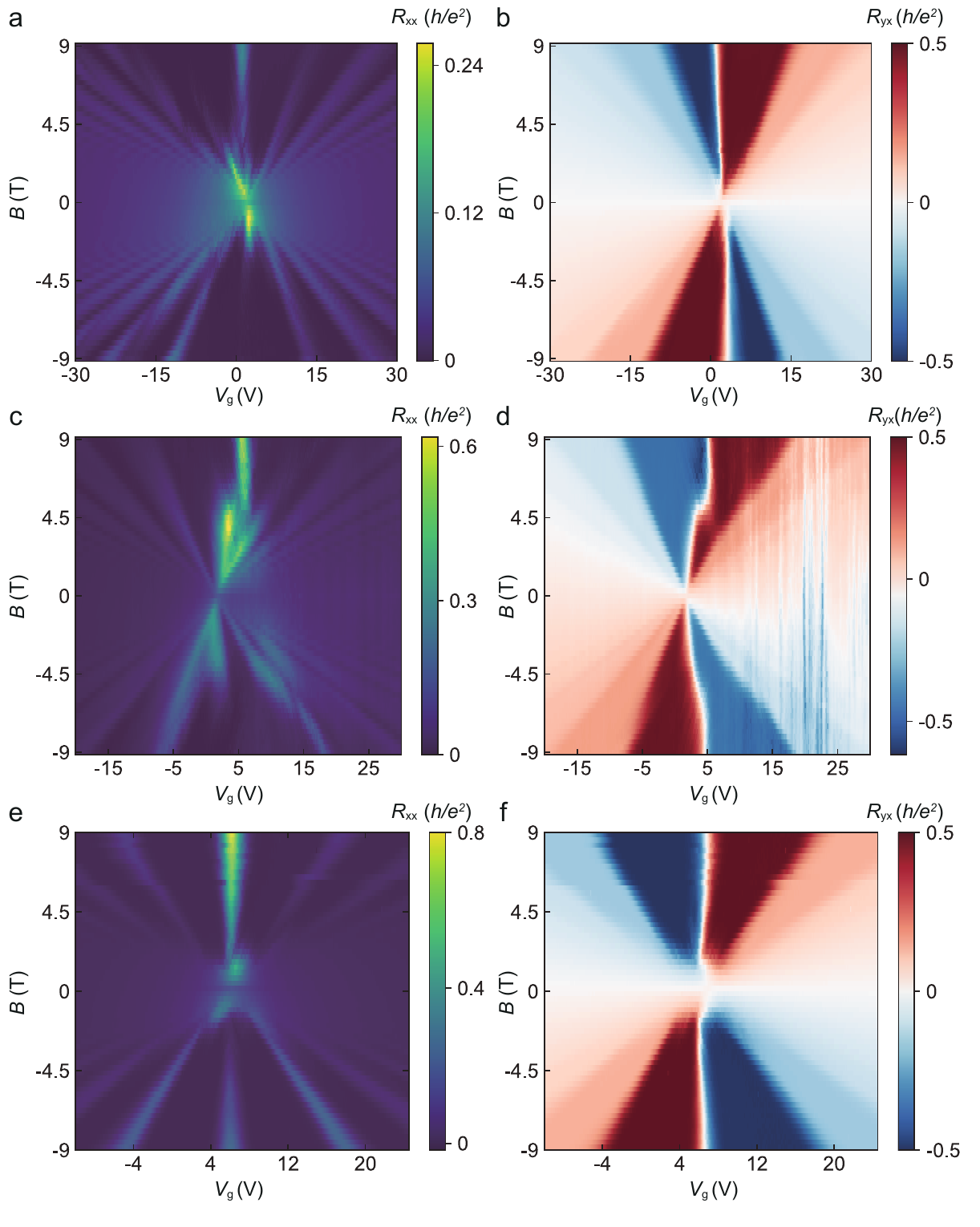}  
\caption{(a, b) Fan diagram for device D2 shows asymmetric quantum Hall effect. Within the range of $-2\,\mathrm{T}$ to $-3\,\mathrm{T}$, the peak of longitudinal resistance at $\nu=0$ drops sharply and decreases to near zero starting at $B = -4$ T. Throughout the entire magnetic field range, asymmetric plateau width in $R_{yx}$ at $\nu=\pm2$ is stably present. (c, d) Fan diagram for device D3. When the negative magnetic field reaches $-7\,\mathrm{T}$, the peak in longitudinal resistance at $\nu=0$ drops to near zero. A shift of the charge neutrality point with magnetic field is also visible in $R_{yx}$. (e, f) Fan diagram for device D4. A pronounced asymmetry in the peak heights at $\nu=0$, $\nu=2$, and $\nu=-2$ is observed under both positive and negative magnetic fields.}
\label{figs_4}
\end{figure*}

Complementing these fixed-field measurements, the full Landau fan diagrams in Fig.~\ref{figs_4} illustrate the stability and field-evolution of these asymmetries. For Device D2 (Fig.~\ref{figs_4} a, b), the asymmetric quantum Hall effect persists robustly across the full magnetic field range ($B = -9$ T to $B = +9$ T). A sharp suppression of the $\nu=0$ $R_{xx}$ peak occurs between $-2\,\mathrm{T}$ and $-3\,\mathrm{T}$, dropping to near zero from $B = -4$ T onward.

Device D3 (Fig.~\ref{figs_4} c, d) exhibits a similar trend, with the $\nu=0$ $R_{xx}$ peak dropping to near zero when the negative magnetic field reaches $-7\,\mathrm{T}$. The fan diagram displays vertical striations in both $R_{xx}$ and $R_{yx}$, most notably within the quantum Hall plateaus, which we attribute to unstable electrical contact at the measurement probes, as we observed a deterioration in electrical contact on these probes during cooldown. The major asymmetric features, particularly the suppression of the $\nu=0$ state, remain clearly resolvable in the figure.

Finally, the Landau fan diagram for Device D4 (Fig.~\ref{figs_4} e, f) clearly demonstrates a pronounced asymmetry in the longitudinal resistance peaks at $\nu = 0$, $\nu = 2$, and $\nu = -2$ under both magnetic field polarities.

This asymmetric behavior was consistently observed across multiple devices, manifested as suppressed $R_{xx}$ peaks at opposite magnetic field polarities and asymmetric Hall plateaus.

\section*{S3: Temperature Dependence}
\label{sec3}
The temperature evolution of the asymmetric quantum Hall effect provides insights into its underlying mechanisms. We systematically investigated this dependence across multiple devices to identify characteristic energy scales and assess the robustness of the observed phenomena.

\begin{figure*}[htbp]
\centering
\includegraphics[width=0.8\textwidth]{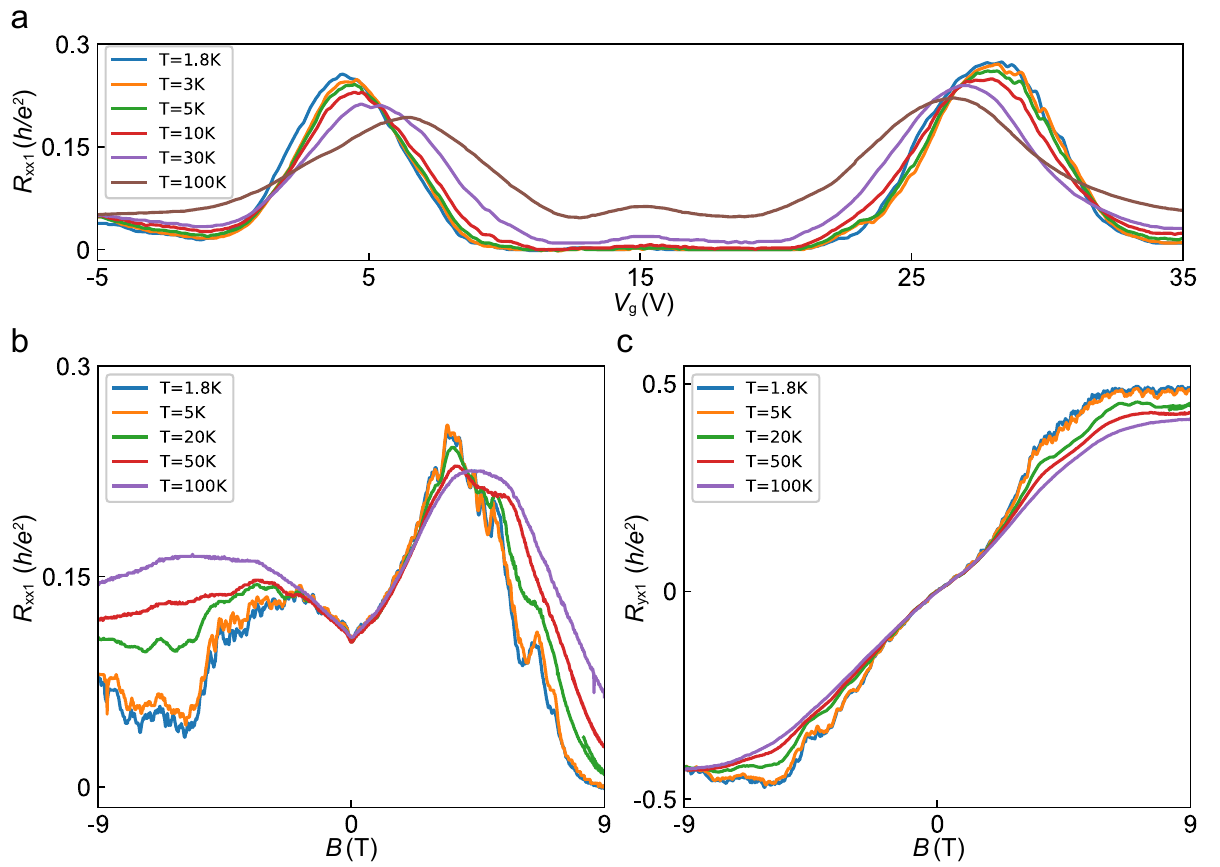}  
\caption{\textbf{Temperature-dependent magnetotransport in Device D1.}
(a) Temperature dependence of $R_{xx1}$ for device D1 at $B = 9\,\mathrm{T}$. With increasing temperature, the peak at $\nu=0$ gradually recovers and increases in height. (b,c) Magnetic field dependence of $R_{xx}$ and $R_{yx}$ for device D1 at $V_g = 20$ V for different temperatures. The asymmetry in $R_{xx}$ is significant across all temperatures and becomes more pronounced at lower temperatures.}
\label{figs_5}
\end{figure*}

The temperature dependence of $R_{xx1}$ for Device D1 at $B = 9\,\mathrm{T}$ is presented in Fig.~\ref{figs_5}(a). As temperature increases, the suppressed $\nu=0$ resistance peak gradually recovers and gains amplitude, indicating thermal activation processes across a finite energy gap. This recovery suggests that the suppression mechanism weakens when thermal energy ($k_BT$) becomes comparable to the characteristic energy scales governing the effect. The energy gaps $E_g$ at $\nu = 0$ were extracted from Arrhenius analysis using the relation $R_{xx} \propto \exp(-E_g / 2k_B T)$, where $k_B$ is the Boltzmann constant and $T$ is the electron temperature. The extracted value of $E_g$ at $\nu = 0$ is $3.99 \,\mathrm{meV}$.

The temperature dependence is further explored through the magnetic field response of $R_{xx}$ and $R_{yx}$ for Device D1 at $V_g = 20\,\mathrm{V}$ in Fig.~\ref{figs_5}(b, c). Remarkably, the transport asymmetry persists across all measured temperatures and becomes increasingly pronounced at lower temperatures. This enhancement at cryogenic conditions suggests the involvement of coherent quantum transport processes that are reinforced when thermal fluctuations are minimized.
\par

\begin{figure*}[htbp]
\centering
\includegraphics[width=0.8\textwidth]{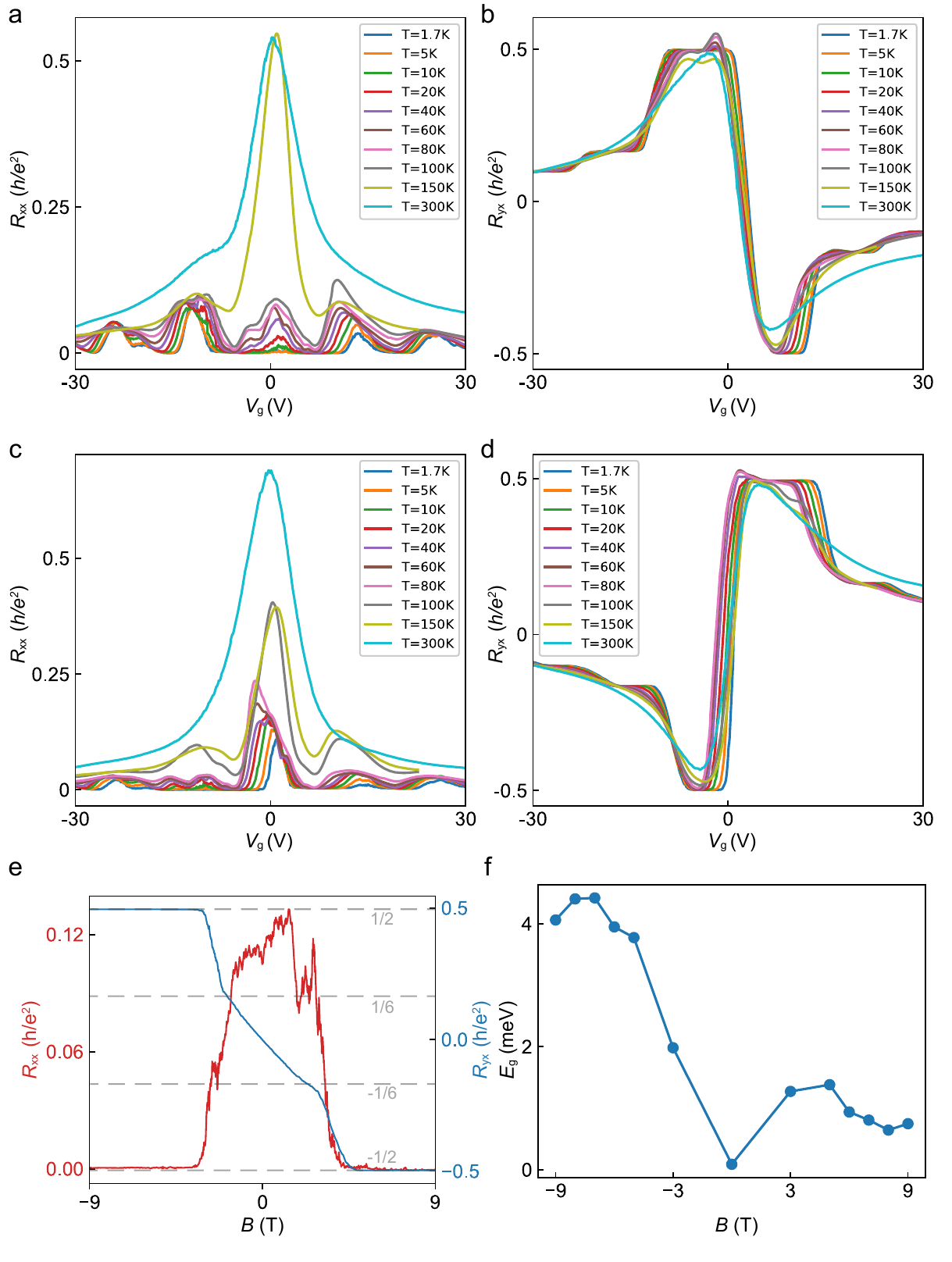}
\caption{
\textbf{Temperature-dependent magnetotransport in Device D2.}
(a, b) Longitudinal and Hall resistances at $B = -9\,\mathrm{T}$, showing a pronounced thermal enhancement of the $\nu=0$ peak between 100--150\,K.
(c, d) Corresponding measurements at $B = +9\,\mathrm{T}$, where the $\nu=0$ peak remains robust across the temperature range.
(e) Magnetic field dependence of $R_{xx}$ and $R_{yx}$ at $T = 1.7\,\mathrm{K}$ and $V_g = -2\,\mathrm{V}$, exhibiting quantized Hall plateaus at $\nu = \pm 2$.
(f) Extracted energy gap $E_g$ at $\nu = 0$ as a function of magnetic field, revealing significant asymmetry between positive and negative fields.
}
\label{figs_6}
\end{figure*}

A detailed temperature-dependent characterization of Device D2 is presented in Fig.~\ref{figs_6}. Under a magnetic field of $B = -9\,\mathrm{T}$, the $\nu = 0$ peak exhibits a progressive enhancement with increasing temperature, with the most pronounced changes occurring between 100–150 K, as shown in Fig.~\ref{figs_6}(a, b). This temperature range likely corresponds to the characteristic energy scale of the suppression mechanism. In contrast, at $B = 9\,\mathrm{T}$, the $\nu = 0$ peak remains relatively stable across the temperature range, showing only a moderate increase at higher temperatures, illustrated in Fig.~\ref{figs_6}(c, d). This distinct behavior under opposite magnetic field polarities underscores the fundamental asymmetry of the quantum Hall response in these heterostructures.

The magnetic field dependence of $R_{xx}$ and $R_{yx}$ for Device D2 was measured at $T = 1.7\,\mathrm{K}$ and $V_g = -2\,\mathrm{V}$, where well-quantized Hall plateaus are observed at $\nu = \pm 2$, as summarized in Fig.~\ref{figs_6}e. To quantitatively evaluate the relevant energy scales, temperature-dependent measurements of $R_{xx}$ were performed at various magnetic fields. The energy gap $E_g$ at $\nu = 0$ was determined at each field through Arrhenius analysis, following the previously described method. The resulting dependence of $E_g$ on magnetic field reveals a pronounced asymmetry between positive and negative fields, presented quantitatively in Fig.~\ref{figs_6}f. This marked gap asymmetry provides clear evidence for energy scales that depend on magnetic field polarity in the graphene/InSe heterostructure.

Overall, the temperature-dependent data indicate that the asymmetric quantum Hall effect is most prominent at low temperatures, yet remains discernible across a broad temperature range. The systematic temperature evolution of the transport characteristics offers important insights for theoretical models of quantum transport mechanisms in graphene/InSe heterostructures.
\par

\section*{S4: Ferroelectricity in Devices of Longitudinal and Hall Resistances}
\label{sec4}

\begin{figure*}[htbp]
\centering
\includegraphics[width=0.8\textwidth]{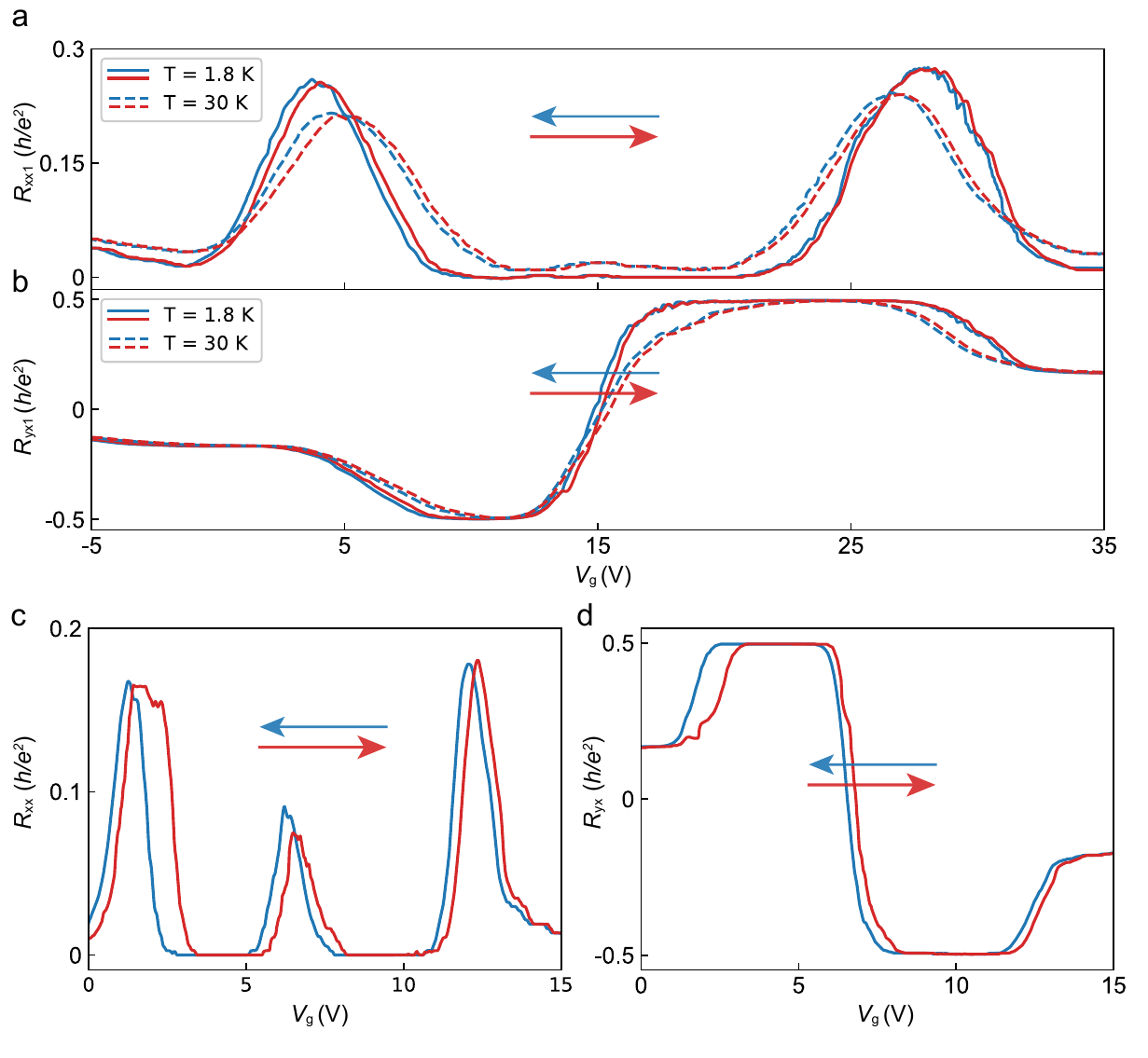}  
\caption{(a) Longitudinal resistance $R_{xx1}$ of device D1 measured at $B = 9$\,T and temperatures $T = 1.8$\,K and 30\,K, showing forward and backward gate sweeps ($V_g$). Arrows indicate the sweep direction. (b) Corresponding $R_{yx1}$ under the same conditions. (c) $R_{xx}$ of device D4 under $B = -4$\,T and $T = 1.7$\,K with bidirectional $V_g$ sweeps. (d) Simultaneously measured Hall resistance $R_{yx}$ of device D4. A small hysteresis on the order of $\sim 1$\,V is observed in all traces, attributed to ferroelectric behavior in the $\gamma$-InSe layer rather than charge trapping in the $\mathrm{SiO_{2}}$ substrate.}
\label{figs_7}
\end{figure*}

A point worth mentioning is that due to $\gamma$-InSe's lack of inversion symmetry, it can host ferroelectricity \cite{Hu2021, Sui2023a}. Gate hysteresis measurements reveal a small, delayed hysteresis on the order of $1\,\mathrm{V}$ in most samples (Fig.~\ref{figs_7}), contrary to the advanced hysteresis typically seen in $\mathrm{SiO_{2}}$ substrates with charge traps. This minor hysteresis likely results from a combined effect of ferroelectric switching in $\gamma$-InSe and the shielding of substrate disorder by the InSe layer. While this small gate voltage shift does not account for the major asymmetry presented earlier, all data discussed herein were acquired with gate traces scanned from $-5\,\mathrm{V}$ to $35\,\mathrm{V}$ to avoid ambiguity.

As shown in Fig.~\ref{figs_7}(a, b), Device D1 exhibits clear hysteresis in both longitudinal ($R_{xx1}$) and Hall ($R_{yx1}$) resistances at $B = 9\,\mathrm{T}$ and temperatures of $1.8\,\mathrm{K}$ and $30\,\mathrm{K}$. The hysteresis, with a magnitude of approximately $1\,\mathrm{V}$, is consistently observed across both temperatures, demonstrating the stable presence of ferroelectric polarization switching over the measured temperature range. This behavior is further reproduced in Device D4 under different conditions ($B = -4\,\mathrm{T}$, $T = 1.7\,\mathrm{K}$; Fig.~\ref{figs_7}(c, d)), confirming the reproducibility of the ferroelectric effect.

The consistent hysteresis observed across devices and conditions suggests ferroelectric polarization switching in the $\gamma$-InSe layer as the underlying mechanism. This ferroelectric behavior contributes to the device characteristics by providing a stable, switchable polarization that modulates the local electrostatic environment without introducing significant disorder.
\par

\section*{S5: Comprehensive Analysis of Disorder Effects on Transport Asymmetry}
\label{sec5}

\begin{figure*}[htbp]
\includegraphics[width=0.4\textwidth]{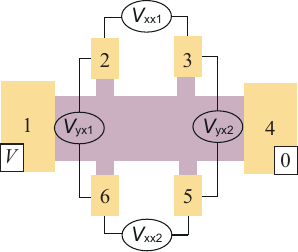}\centering
\caption{A schematic plot of the six-terminal measurement setups.}
\label{figs8}
\end{figure*}

We use the Landauer-B\"{u}ttiker formalism to study the electron transport of the graphene/InSe heterostructures. For a multi-lead configuration, the net current flowing in a given lead \( p \) is provided by~\cite{sdatta_book_1997}
\begin{equation}
    \begin{gathered}
    I_p=\sum_q (G_{qp}V_p-G_{pq}V_q)
    \end{gathered}\label{eq_method_1}
\end{equation}
with \( V_p \) and \( V_q \) being the voltage of leads \( p \) and \( q \), respectively. \( G_{pq} \) is the conductance coefficient from lead \( q \) to \( p \) and reads~\cite{palee_prl_1981}
\begin{equation}
    \begin{gathered}
    G_{pq}=\dfrac{e^2}{h}\text{Tr}[\Gamma_p G^{\text{R}} \Gamma_q G^{\text{A}}]. 
    \end{gathered}\label{eq_method_2}
\end{equation}
In Eq. (4), $ G^{\mathrm{R}(\mathrm{A})}=[E-H-\Sigma^{\mathrm{R}(\mathrm{A})}]^{-1} $ is the retarded (advanced) Green's function with $ \Sigma^{\mathrm{R}(\mathrm{A})} $ being the retarded (advanced) self-energy and $ \Gamma_{p(q)}=i(\Sigma^{\mathrm{R}}_{p(q)}-\Sigma^{\mathrm{A}}_{p(q)}) $. The experimental setups are six-terminal configurations, as shown in Fig. 5, with the source and drain being terminals 1 and 4, and the leads being terminals 2, 3, 5, and 6. Current injection occurs between terminal 1 (source) and terminal 4 (drain), satisfying $ I_{4}=-I_{1} $, while the leads being terminals 2, 3, 5, and 6 maintaining zero net current flow ($ I_{2}=I_{3}=I_{5}=I_{6}=0 $). In terms of matrices, $ \hat{I}=\hat{G}\hat{V} $, i.e.,
\begin{equation}
    \begin{gathered}
    \begin{bmatrix}
    I_1 \\
    I_2 \\
    I_3 \\
    I_4 \\
    I_5 \\
    I_6
    \end{bmatrix}=    
    \begin{bmatrix}
    I_1 \\
    0 \\
    0 \\
    -I_1 \\
    0 \\
    0
    \end{bmatrix}=
    \begin{bmatrix}
    \sum_p G_{1p} & -G_{12} &  -G_{13} & -G_{14} & -G_{15} & -G_{16} \\
    -G_{21} & \sum_p G_{2p} & -G_{23} & -G_{24} & -G_{25} & -G_{26}  \\
    -G_{31} & -G_{32} & \sum_p G_{3p} & -G_{34} & -G_{35} & -G_{36}  \\
    -G_{41} & -G_{42} & -G_{43} & \sum_p G_{4p} & -G_{45} & -G_{46}  \\
    -G_{51} & -G_{52} & -G_{53} & -G_{54} & \sum_p G_{5p} & -G_{56}  \\
    -G_{61} & -G_{62} & -G_{63} & -G_{64} & -G_{65} & \sum_p G_{6p}  \\
    \end{bmatrix}
    \begin{bmatrix}
    V_1 \\
    V_2 \\
    V_3 \\
    V_4 \\
    V_5 \\
    V_6
    \end{bmatrix}.
    \end{gathered}\label{eq_method_3}
\end{equation}
Set $V_4=0$:
\begin{equation}
    \begin{gathered}
    \begin{bmatrix}
    I_1 \\
    0 \\
    0 \\
    0 \\
    0 
    \end{bmatrix}=
    \begin{bmatrix}
    \sum_p G_{1p} & -G_{12} &  -G_{13}  & -G_{15} & -G_{16} \\
    -G_{21} & \sum_p G_{2p} & -G_{23}  & -G_{25} & -G_{26} \\
    -G_{31} & -G_{32} & \sum_p G_{3p}  & -G_{35} & -G_{36} \\
    -G_{51} & -G_{52} & -G_{53}  & \sum_pG_{5p} & -G_{56} \\
    -G_{61} & -G_{62} & -G_{63}  & -G_{65} & \sum_p G_{6p} 
    \end{bmatrix}
    \begin{bmatrix}
    V_1 \\
    V_2 \\
    V_3 \\
    V_5 \\
    V_6 
    \end{bmatrix}. 
    \end{gathered}\label{eq_method_4}
\end{equation}
The resistance $R_L$ and the Hall resistance $R_H$ can be calculated by
\begin{equation}
    \begin{gathered}
    R_L=\dfrac{V_3-V_2}{I_1}=R_{31}-R_{21},R_H=\dfrac{V_6-V_2}{I_1}=R_{61}-R_{21}.
    \end{gathered}\label{eq_method_5}
\end{equation}
Here, \( R_{ij} \) is the element of \( \hat{R}=\hat{G}^{-1} \) if \( \det [\hat{G}]\neq 0 \). We apply the tight-binding model to the setups shown in Fig.~\ref{figs8} and calculate the conductance matrix \( \hat{G} \) by using Eq.~\eqref{eq_method_2}. Then, the resistance and the Hall resistance are calculated using Eq.~\eqref{eq_method_5}, which can be directly compared with experimental data. Numerically, we utilize the well-established Python package Kwant to construct the tight-binding model and calculate the conductance coefficient \( G_{pq} \)~\cite{kwant}. 

In a clean sample under a moderate magnetic field, which corresponds to a pure graphene subject to a perpendicular magnetic field, we obtain the correct sequence of quantum Hall plateaux in the transverse resistance with the corresponding transitional peaks in the longitudinal resistance, validating our numerical method. Moreover, the longitudinal resistance is symmetric between the two sides of the sample and in opposite magnetic field directions. 

Based on this validated numerical framework, we first rule out purely random disorder as the origin of the systematic asymmetry observed at $\nu = 0$. Our experimental observation of a vanishing longitudinal resistance on one side of the sample is inconsistent with a completely random disorder landscape, which would typically produce multiple electron and hole puddles and result in a finite longitudinal voltage drop. This key discrepancy indicates that the monotonic density profile required by our toy model cannot arise from random disorder alone, but must instead result from a gradient mechanism.

\begin{figure*}[htbp]
\centering
\includegraphics[width=0.4\textwidth]{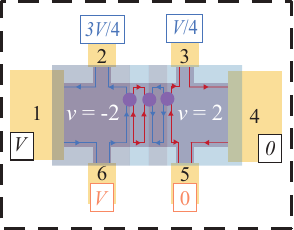}\centering
\caption{
\textbf{Interior edge channel equilibration in a disordered landscape with multiple puddles.} Schematic illustration of a scenario where disorder creates multiple electron (blue) and hole (red) puddles, dividing the sample into several distinct regions of different local filling factors. Edge currents flow along the boundaries between these puddles. In such a configuration, a finite voltage develops between adjacent voltage probes (e.g., Lead 2 and 3) due to the complex percolation paths of chiral edge states navigating the disordered potential landscape.
}
\label{figs_9}
\end{figure*}

To thoroughly investigate the impact of disorder on the observed transport asymmetry, we conducted an extensive series of numerical simulations beyond the specific cases presented in the main text. Our goal was to systematically explore how the interplay between disorder strength $W$ and correlation length $\xi$ influences the longitudinal and Hall resistances. We performed a broad parameter sweep, simulating numerous disorder realizations across a wide range of $W$ (from 0.01 to 0.1) and $\xi$ (from $10a$ to $200a$). This comprehensive computational study confirms that while disorder-induced asymmetry is a common feature, its specific manifestation is highly sensitive to the microscopic characteristics of the disorder potential.

\begin{figure*}[htbp]
\centering
\includegraphics[width=0.8\textwidth]{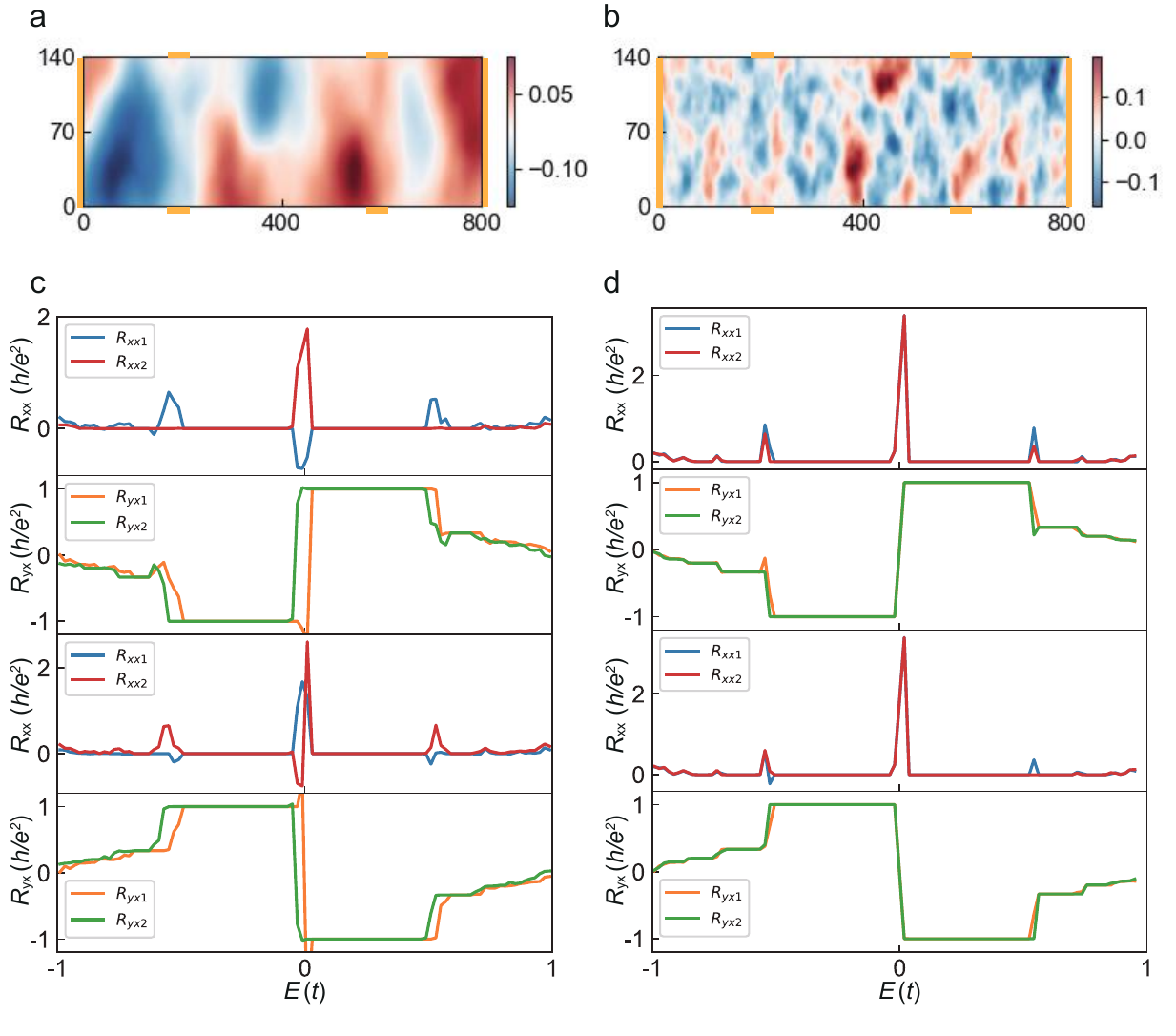}
\caption{
\textbf{Impact of disorder characteristics on potential landscape and transport properties.} (a) Spatial distribution of the disorder potential for parameters $(W=0.02, \xi=100a)$. (b) Spatial distribution for parameters $(W=0.1, \xi=10a)$. (c) Longitudinal ($R_{xx}$) and Hall ($R_{yx}$) resistances as a function of Fermi energy $E$ for the system in (a). (d) Corresponding $R_{xx}$ and $R_{yx}$ for the system in (b). The distinct disorder landscapes lead to markedly different evolutions of the transport coefficients.
}
\label{figs_10}
\end{figure*}

The results of this vast parameter exploration revealed that the emergence of significant asymmetry is not contingent on a single specific disorder configuration but is a robust phenomenon across many different realizations, provided the disorder strength and correlation length fall within an appropriate range. Fig.~\ref{figs_10} presents two more representative examples from our large data set that exemplify the typical outcomes of these simulations.

Our extensive simulations indicate that the correlation length $\xi$ plays a particularly crucial role in determining the transport characteristics. For small $\xi$ (short-range disorder), the potential landscape comprises numerous small, isolated electron and hole puddles. In this regime, while some asymmetry can be observed, the overall transport behavior tends to be more symmetric compared to cases with longer correlation lengths. As $\xi$ increases to approximately $150a$ (as shown in Fig.~\ref{figs_10}), the disorder potential develops longer-range fluctuations, creating more extended and connected puddle structures. This intermediate correlation length regime proves most effective in generating the pronounced asymmetry seen in our experiments. For very large $\xi$ values approaching the device size, the disorder potential effectively morphs into a smooth gradient similar to that discussed in the main text.

The disorder strength $W$ similarly influences the degree of asymmetry. At very low $W$ values, the disorder potential is too weak to significantly disrupt the native transport properties of the system. As $W$ increases to the range $0.02-0.04$, we observe the strongest asymmetry effects, with the specific value depending on the correlation length $\xi$ (Fig.~\ref{figs_10} a,c). Beyond this optimal range, excessive disorder strength tends to overwhelm the quantum Hall effect altogether, suppressing both the quantized plateaus and the transitional peaks (Fig.~\ref{figs_10} b,d).

\begin{figure*}[htbp]
\centering
\includegraphics[width=0.6\textwidth]{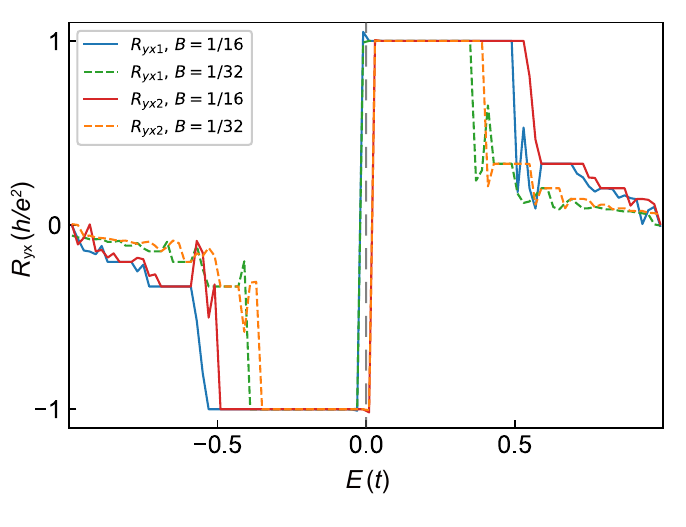}
\caption{
\textbf{Numerical simulation of the $R_H = 0$ line shift.}
The calculated variations of $R_{yx1}$ and $R_{yx2}$ with energy $E$ under different magnetic fields are shown. The gray dashed line indicates the position of $E = 0$.}
\label{figs_11}
\end{figure*}

The offset of the charge neutrality point (CNP) also emerges in systems with a substantial disorder potential. Fig.~\ref{figs_11} shows the calculated $R_{yx}$ under the chemical potential profile of Fig. 4b in the main text. At finite magnetic fields, the gate voltage at which $R_{yx1}$ changes sign shifts toward lower energies, while that for $R_{yx2}$ shifts toward higher energies.

In conclusion, our comprehensive computational investigation demonstrates that transport asymmetry represents a generic feature of quantum Hall systems with underlying disorder. The specific nature of this asymmetry---including the relative amplitudes of different peaks and their evolution with magnetic field---varies considerably across different disorder realizations and parameters. However, the complete suppression of a longitudinal resistance peak, as observed in some devices, appears to require a more specific potential landscape than that provided by typical random disorder. This suggests that a monotonic density gradient, rather than random disorder alone, is likely responsible for the most dramatic asymmetric effects reported in the main text.
\par\quad\par

\clearpage

\bibliography{mainrefs-V18}

\clearpage